\documentclass[11pt,twoside]{article}
\usepackage{FUSE2004}
\usepackage{natbib}

\usepackage{epsf}
\usepackage{psfig}
\usepackage{lscape}

\begin{document}
\title{Missing Baryons in the Warm-Hot Intergalactic Medium}
\author{Todd M. Tripp}
\affil{University of Massachusetts, Dept. of Astronomy, 710 N. Pleasant St., Amherst, MA 01003}
\author{David V. Bowen}
\affil{Princeton University Observatory, Peyton Hall, Ivy Lane, Princeton, NJ 08544}
\author{Kenneth R. Sembach}
\affil{Space Telescope Science Institute, 3700 San Martin Dr., Baltimore, MD 21218}
\author{Edward B. Jenkins}
\affil{Princeton University Observatory, Peyton Hall, Ivy Lane, Princeton, NJ 08544}
\author{Blair D. Savage}
\affil{University of Wisconsin, Dept. of Astronomy, 475 N. Charter St., Madison, WI 53706}
\author{Philipp Richter}
\affil{Institut f\"{u}r Astrophysik und Extraterrestrische Forschung, Universit\"{a}t Bonn, Aug dem H\"{u}gel 71, 53121 Bonn, Germany}

\begin{abstract}
This review briefly discusses the use of UV absorption lines in the
spectra of low-redshift QSOs for the study of the physical conditions,
metallicity, and baryonic content of the low$-z$ IGM, with emphasis on
implications for the missing baryons problem. Current results on the
statistics and baryonic content of intervening, low$-z$ \ion{O}{6} and
Ly$\alpha$ absorption-line systems are presented with some comments on
overlap between these two classes of absorbers and consequent baryon
double-counting problems. From observations of a sample of 16 QSOs
observed with the E140M echelle mode of STIS, we find 44 intervening
\ion{O}{6} absorbers and 14 associated \ion{O}{6} systems (i.e,
systems with $z_{\rm abs} \approx z_{\rm em}$). This sample implies
that the number of intervening \ion{O}{6} absorbers per unit redshift
is $dN/dz$(\ion{O}{6}) = $23\pm 4$ for lines with rest equivalent
width $>$ 30 m\AA . The intervening \ion{O}{6} systems contain at
least 7\% of the baryons if their typical metallicity is 1/10 solar
and the \ion{O}{6} ion fraction is $\leq$0.2.  This finding is
consistent with predictions made by cosmological simulations of
large-scale structure growth. Recently, a population of remarkably
broad Ly$\alpha$ lines have been recognized in low$-z$ quasar
spectra. If the breadth of these Ly$\alpha$ features is predominantly
due to thermal motions, then these \ion{H}{1} absorbers likely harbor
an important fraction of the baryons.  We present and discuss some
examples of the broad Ly$\alpha$ absorbers. Finally, we summarize some
findings on the relationships between \ion{O}{6} absorbers and nearby
galaxies/large-scale structures.
\end{abstract}

\section{Introduction}

It is important to understand the distribution, physical conditions,
and metallicity of the baryons in the Universe, both as a function of
redshift (i.e., time) and as a function of environmental factors
(e.g., proximity to galaxies, galaxy groups/clusters, or voids).
Ordinary baryonic matter plays a fundamental role in galaxy formation
and evolution, and the state of the baryons often provides a stringent
test of cosmological theories. Measurements of deuterium abundances in
low-metallicity QSO absorption systems (e.g., Burles \& Tytler 1998;
O'Meara et al. 2001; Sembach et al. 2004a) and observations of the
cosmic microwave background (e.g., Spergel et al. 2003) provide good
estimations of the {\it total} quantity of baryons expected to be
found in the Universe; expressed as the usual units of the closure
density, these measurements indicate that $\Omega _{b} h_{75}^{2} =
0.04$.  At high redshifts, it appears that the matter in the
Ly$\alpha$ forest can account for the majority of the expected baryons
(Weinberg et al. 1997; Rauch et al. 1997; Schaye 2001), but attempts
to inventory the baryons in the nearby Universe generally fail to find
enough ordinary matter in the form of stars, hot (X-ray emitting) gas
in clusters, and cold gas (see Fukugita et al. 1998, and references
therein); the sum of these well-observed baryon reservoirs appears to
be at least a factor of 2 lower than the predicted amount at the
present epoch.

\begin{figure}
\plottwo{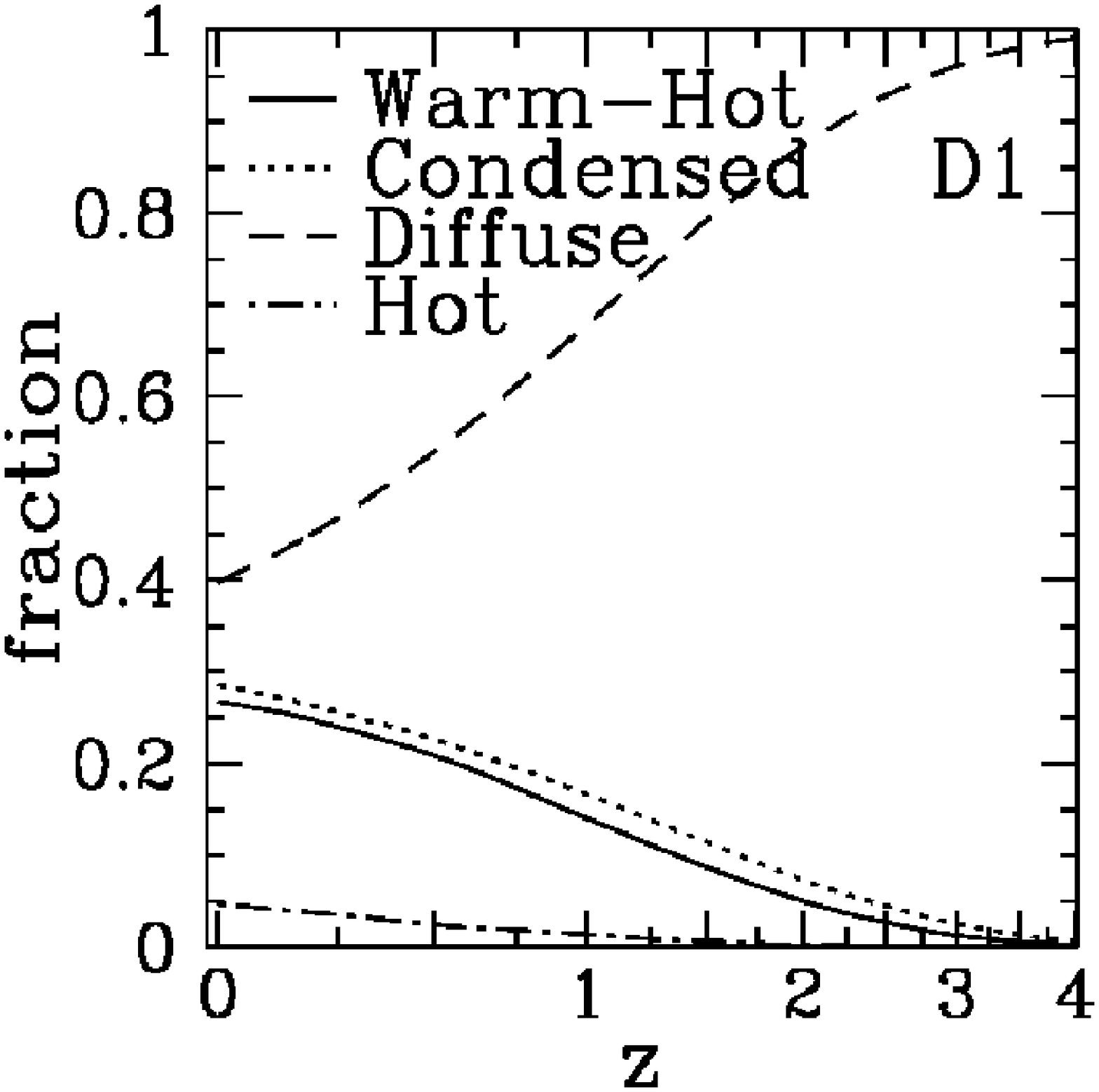}{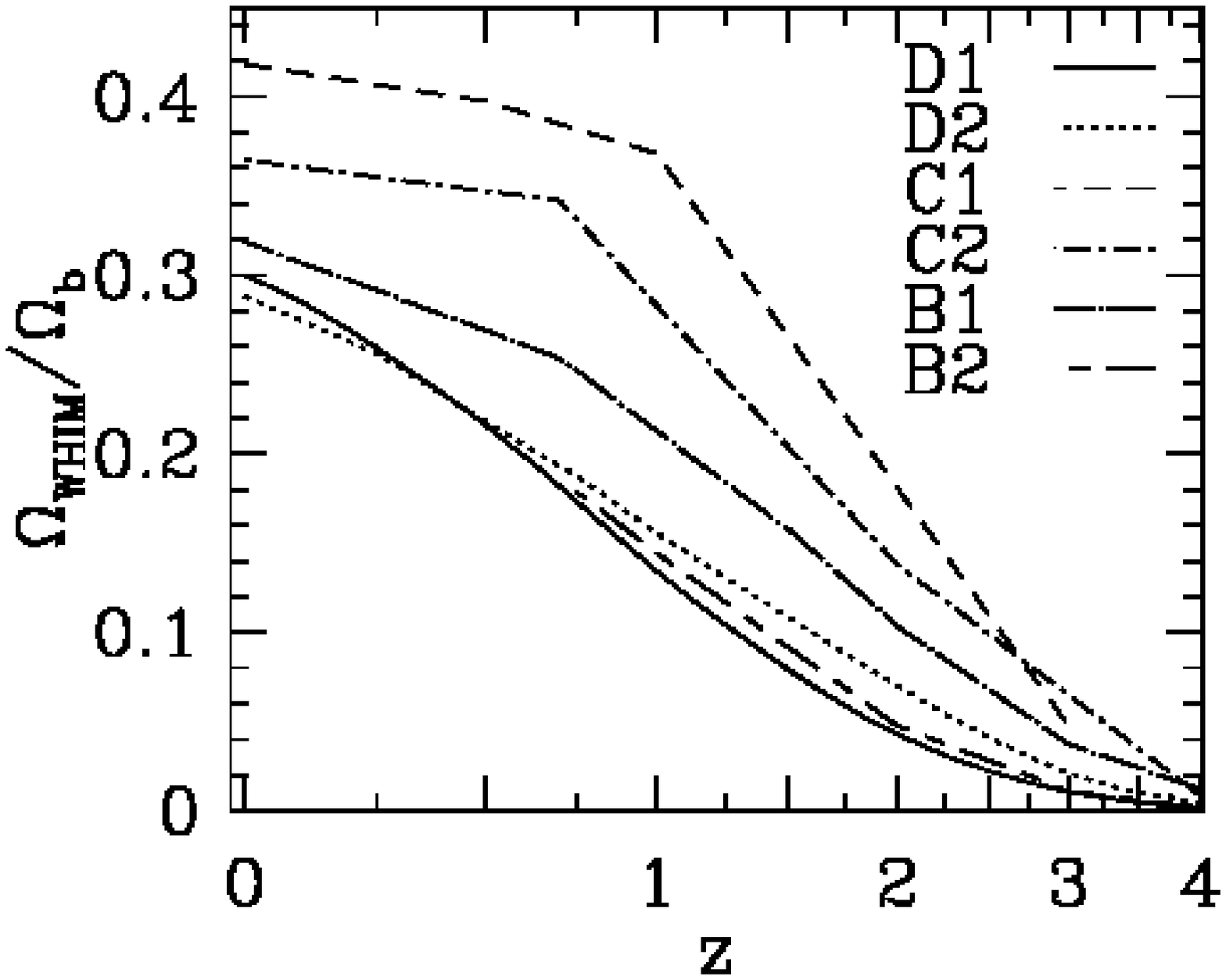}
\caption{Predicted distributions of baryons from CDM cosmological
simulations, reproduced from Dav\'{e} et al. (2001). {\bf Left panel:}
The mass fractions as a function of redshift of baryons in the form of
relatively cool, photoionized Ly$\alpha$ clouds (``diffuse'', dashed
curve), stars in galaxies (``condensed'', dotted curve), shock-heated
intergalactic gas at $10^{5} - 10^{7}$ K (``warm-hot'', solid curve),
and hotter gas in clusters that produces detectable X-ray emission
(``hot'', dash-dot curve). {\bf Right panel:} The predicted fraction
of the baryons in the warm-hot gas vs. redshift from six different
hydrodynamic CDM simulations that use a variety of numerical methods
and have significantly different resolutions, box sizes, etc (see
Dav\'{e} et al.).\label{models}}
\end{figure}

A variety of solutions to this ``missing baryons'' problem have been
proposed, and considerable observational effort is currently being
invested to test these ideas.  One possible solution has emerged from
hydrodynamic simulations of cosmological structure growth. These
simulations predict that at high redshifts, all of the baryons were
indeed in the cool, photoionized Ly$\alpha$ forest, but as time
passed, much of that matter collapsed and turned into galaxies, and
much of the Ly$\alpha$ forest that didn't turn into stars was shock
heated into the $10^{5} - 10^{7}$ K temperature range. In these
models, at the present time the $10^{5} - 10^{7}$ K baryons remain
located in the low-density intergalactic gas, the so-called ``warm-hot
intergalactic medium'' (WHIM or WHIGM).\footnote{In this paper we use
the WHIGM (pronounced ``wiggum'') acronym because ``WHIM'' can be
confused with ``WIM'', the well-known and widely-studied warm ionized
medium phase of galactic interstellar media.}  As shown in
Figure~\ref{models}, many different CDM cosmological simulations
robustly make the same prediction with regard to baryonic content of
the WHIGM: currently, roughly 1/3 of the baryons are in stars in
galaxies, 1/3 are in the shock-heated WHIGM, and 1/3 remain in the
cool photoionized Ly$\alpha$ forest (Cen \& Ostriker 1999; Dav\'{e} et
al. 2001, and references therein). Since the WHIGM result reproducibly
emerges from a diverse set of cosmological simulations, observational
evidence that a significant portion of the baryons are located in the
low$-z$ WHIGM would therefore not only solve the long-standing missing
baryons problem and provide important constraints for galaxy evolution
modeling, this would also indicate that at least in some regards, the
cosmological simulations are able to capture enough of the IGM physics
adequately to make correct predictions.

\section{Statistics and Baryonic Content of Low-z \ion{O}{6} Absorbers}

\begin{figure}
\plotone{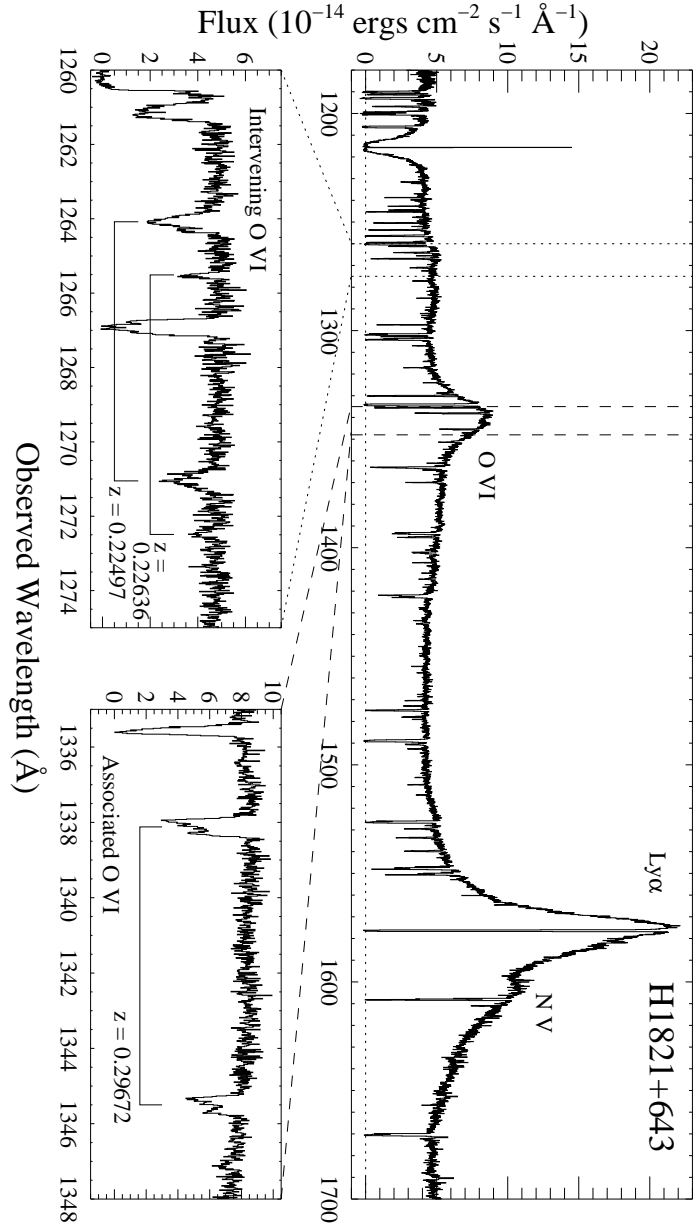}
\caption{{\bf Upper panel:} STIS E140M echelle spectrum of H1821+643
with the Ly$\alpha$, \ion{N}{5}, and \ion{O}{6} emission lines of the
QSO marked. {\bf Lower panels:} examples of the \ion{O}{6} $\lambda
\lambda 1031.93, 1037.62$ doublet in absorption including (lower left)
the intervening absorption systems at $z_{\rm abs}$ = 0.22497 and
0.22636, and (lower right) the multicomponent associated absorber
(i.e., $z_{\rm abs} \approx z_{\rm QSO}$) at $z_{\rm abs}$ = 0.29672.
The upper panel is binned, but the lower panels show the spectrum at
full, unbinned resolution (FWHM = 7 km s$^{-1}$).\label{stisexample}}
\end{figure}

\begin{figure}
\plotone{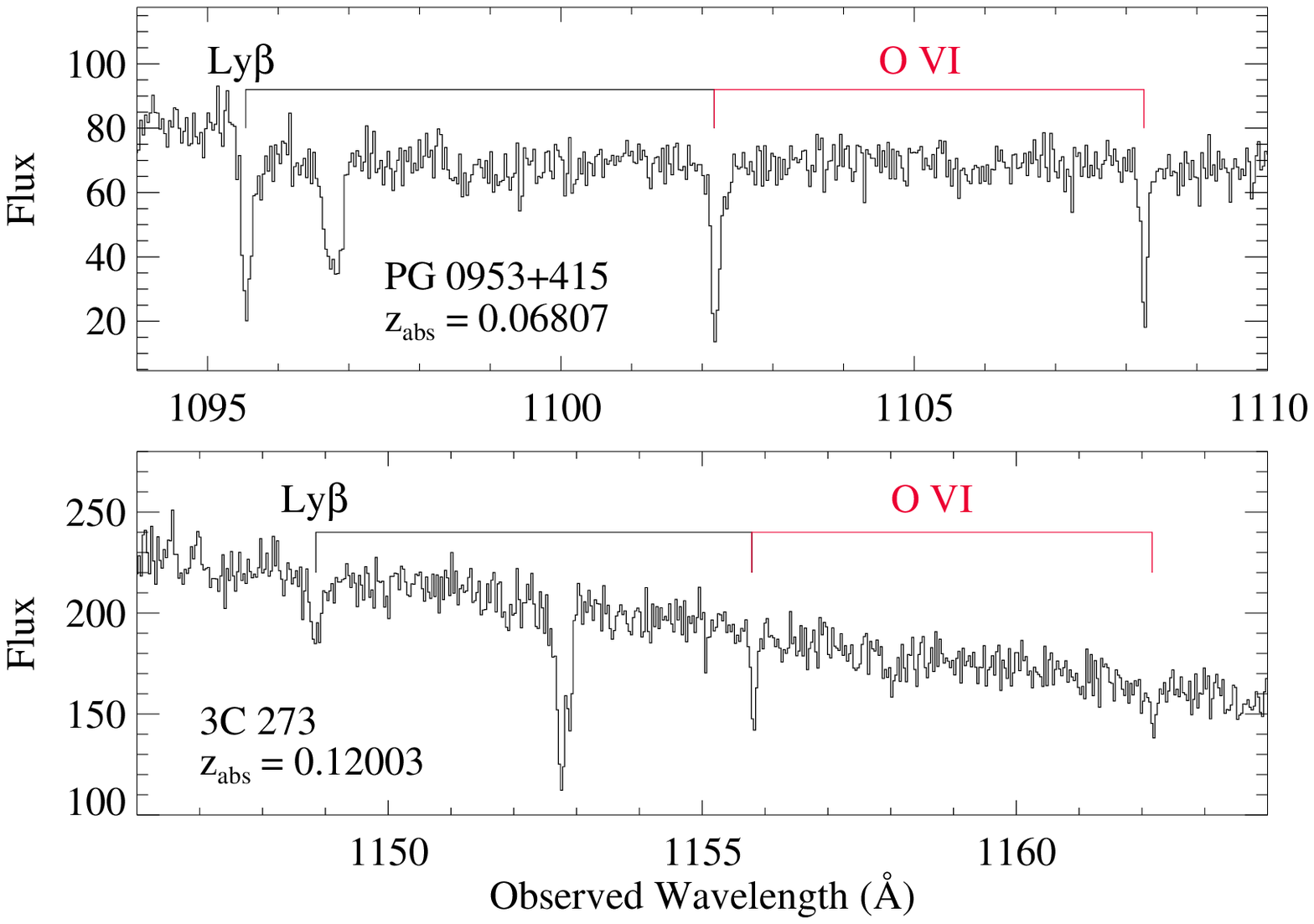}
\caption{Examples of intervening \ion{O}{6} absorbers detected with
{\it FUSE} in the spectra of PG0953+415 (upper panel, see Savage et
al. 2002) and 3C 273 (lower panel, see Sembach et al. 2001) including
the Ly$\beta$ and \ion{O}{6} $\lambda \lambda$1031.93, 1037.62 lines
at $z_{\rm abs}$ = 0.06807 (PG0953+415) and $z_{\rm abs}$ = 0.12003
(3C 273).\label{fuseexample}}
\end{figure}

Due to its low density and temperature range, the WHIGM is expected to
be exceedingly difficult to detect in emission.  Absorption
spectroscopy provides a much more sensitive means to study low-density
gas, and in the near term this technique is likely to provide the
primary observational constraints on the low-redshift WHIGM. In the
ultraviolet, the resonance line doublet of Li-like \ion{O}{6} at
1031.93 and 1037.62 \AA\ is particularly useful for WHIGM observations
because (1) in collisional ionization eq., \ion{O}{6} peaks in
abundance at log $T \approx$ 5.5, (2) oxygen is the most abundant
metal, and (3) the lines have relatively large $f-$values. 

To search for low$-z$ \ion{O}{6} absorbers and the missing baryons, we
have been observing low-redshift QSOs for several years with both the
Space Telescope Imaging Spectrograph (STIS) and the {\it Far
Ultraviolet Spectroscopic Explorer (FUSE)}. These observations have
revealed a large number of intervening and associated (i.e., $z_{\rm
abs} \approx z_{\rm QSO}$) \ion{O}{6} systems.  Examples of the
detected \ion{O}{6} lines are shown in Figures~\ref{stisexample} and
\ref{fuseexample}. The first QSO that we observed for this purpose
(H1821+643, shown in Figure~\ref{stisexample}) unveiled a surprisingly
large number of intervening systems (Tripp, Savage, \& Jenkins 2000;
Oegerle et al. 2000). The detection of six intervening \ion{O}{6}
absorbers toward H1821+643 indicated an incidence of \ion{O}{6}
systems per unit redshift $dN/dz \approx$ 50 for lines with rest
equivalent width $W_{\rm r} \geq 30$ m\AA .  Observations of other
sight lines confirmed the redshift density of \ion{O}{6} lines at $z <
0.6$ (Tripp \& Savage 2000; Savage et al. 2002; Richter et al. 2004,
Sembach et al. 2004b; Prochaska et al. 2004).  To estimate the
baryonic content of the \ion{O}{6} systems, information (or
assumptions) about the gas metallicity (O/H) and \ion{O}{6} ionization
fraction $f$(\ion{O}{6}) are required.  However, since $\Omega_{\rm
b}$(\ion{O}{6}) $\propto f$(\ion{O}{6})$^{-1}$(O/H)$^{-1}$, it is
possible to place useful lower limits on $\Omega_{\rm b}$(\ion{O}{6})
by assuming the maximum plausible values for $f$(\ion{O}{6}) and
(O/H). For \ion{O}{6}, the ion fraction rarely exceeds 0.2 (Tripp \&
Savage 2000), but it is less clear what the maximum (or even the
typical) metallicity might be. Assuming (O/H) = 0.1 (O/H)$_{\odot}$
and $f$(\ion{O}{6}) $\leq$ 0.2, we find that low$-z$ \ion{O}{6}
absorbers harbor 5\% or more of the baryons.

The baryonic content and $dN/dz$ of the low$-z$ \ion{O}{6} lines
derived from the first observations are consistent with the
predictions from a variety of cosmological simulations (Cen et
al. 2001, Fang \& Bryan 2001; Chen et al. 2003). However, the first
observational papers suffered considerably from small-sample
statistics.  To rectify this problem, we have acquired additional
observations of low$-z$ QSOs with the E140M echelle mode of
STIS. Combined with additional E140M data from the {\it HST} archive,
we now have a sample of 16 QSOs with $0.1583 \leq z_{\rm QSO} \leq
0.5726$. At the time of this writing, we have found 44 intervening
\ion{O}{6} absorbers and 14 associated \ion{O}{6} systems in these
spectra.  We have carefully assessed the total redshift path probed by
this sample (e.g., we have corrected the path to account for regions
that are blocked by strong ISM or unrelated extragalactic lines, and
we have accounted for S/N variations vs. $\lambda$ for each sight
line). With this substantially larger \ion{O}{6} sample, we find for
the intervening systems $dN/dz$(\ion{O}{6}) = $23\pm 4$ for lines with
rest equivalent width $>$ 30 m\AA . Again assuming (O/H) = 0.1
(O/H)$_{\odot}$ and $f$(\ion{O}{6}) $< 0.2$, we obtain $\Omega_{\rm
b}$(\ion{O}{6}) = 0.0027 or $\Omega_{\rm b}$(\ion{O}{6})/$\Omega_{\rm
b}$(total) $\geq$ 0.068, i.e., 7\% of the baryons. 

It is important to note that we have used the recent revision of the
solar oxygen abundance reported by Allende Prieto et al. (2001) to
determine the value of (O/H) = 1/10 (O/H)$_{\odot}$ for this
calculation of $\Omega_{\rm b}$(O~VI). Our previous papers used
(O/H)$_{\odot}$ from Grevesse \& coworkers (e.g., Grevesse et
al. 1996); the Allende Prieto et al. value for (O/H)$_{\odot}$ is
substantially lower. Since $\Omega_{\rm b}$(\ion{O}{6}) $\propto$
(O/H)$^{-1}$, the substantial reduction in the solar oxgyen abundance
reported by Allende Prieto et al. leads to an increase in
$\Omega_{\rm b}$(O~VI).  Taking into account the revision of
(O/H)$_{\odot}$, the new constraints on $\Omega_{\rm b}$(\ion{O}{6})
are in excellent agreement with the earlier findings.  Our sample
applies to systems with $z_{\rm abs} >$ 0.12; we note that a largely
independent sample of \ion{O}{6} systems at $z_{\rm abs} <$ 0.15
compiled by Danforth \& Shull (2004) yields very similar values for
$dN/dz$ and $\Omega_{\rm b}$(\ion{O}{6}).  We are continuing to add
data to our sample, mainly by adding {\it FUSE} observations of the
same sight lines, which provide information about \ion{O}{6} lines at
$z_{\rm abs} <$ 0.13.

\section{Ionization and Metallicity of the \ion{O}{6} Absorbers}

But what about the ionization and metallicity of the \ion{O}{6}
systems? Is 1/10 solar metallicity an appropriate assumption, and
moreover are the absorbers collisionally ionized as expected for the
(shocked) WHIGM? Low$-z$ absorption systems appear to have a wide
range of metallicities: some show very low abundances (e.g., Tripp et
al. 2002; 2004) while others have attained $Z \geq 0.5 Z_{\odot}$
(e.g., Savage et al. 2002, 2004; Jenkins et al. 2005). Likewise, some
\ion{O}{6} systems appear to be photoionized by the UV background
(e.g., Savage et al. 2002), but some \ion{O}{6} lines arise in
collisionally ionized, hot gas (e.g., Tripp et al. 2001; Savage et
al. 2004). Moreover, there is undeniable evidence that many \ion{O}{6}
systems are multiphase entities (Tripp et al. 2000, 2001; Shull et
al. 2003; Richter et al. 2004; Sembach et al. 2004b; Savage et
al. 2004; Tumlinson et al. 2004). In the near term, the multiphase
nature of the \ion{O}{6} systems complicates the interpretation of the
absorbers, but ultimately the rich information in these cases will
likely be very valuable for understanding their detailed nature. 

\section{Baryonic Content of Low-z Ly$\alpha$ Lines and Double Counting}

What about baryons in the cool, photoionized Ly$\alpha$ clouds at low
redshifts? There is good agreement regarding the statistics (e.g.,
$dN/dz$) of low$-z$ Ly$\alpha$ lines (compare Tripp et al. 1998;
Penton et al. 2000a, 2004; Richter et al. 2004; Sembach et
al. 2004b). To estimate the baryonic content of the Ly$\alpha$
absorbers, very large ionization corrections must be applied, but at
least no metallicity information is needed. Estimations of the total
baryonic mass in the low$-z$ Ly$\alpha$ forest (e.g., Shull, Stocke,
\& Penton 1996; Penton et al. 2000a,2004) indicate that $\sim$30\% of
the baryons are in these systems, in agreement with the cosmological
simulations.

There is certainly overlap between \ion{O}{6} and Ly$\alpha$
absorbers, and this raises a concern about how to fold the
contributions of the \ion{O}{6} and Ly$\alpha$ absorbers into the
overall inventory of baryons: if some of the \ion{O}{6} systems are
photoionized and have corresponding \ion{H}{1} lines (e.g., Savage et
al. 2002), then these baryons could be double-counted.  However, it is
probably also true that some of the Ly$\alpha$ lines which are assumed
to arise in cool, photoionized gas actually originate in hot gas, so
the double-counting problem operates in both directions.  And, it is
possible (even likely) that many systems contain a mix of hot and cool
gas. We note that photoionized \ion{O}{6} systems are expected in the
cosmological models (Cen et al. 2001; Fang \& Bryan 2001). Solution of
this problem will require large samples of absorbers with good-quality
line measurements.

\section{Broad Ly$\alpha$ Lines}

Several papers have presented evidence of broad Ly$\alpha$ lines with
$b-$values in excess of 40 km s$^{-1}$ (e.g., Tripp et al. 2001;
Bowen, Pettini, \& Blades 2002; Richter et al. 2004; Sembach et
al. 2004b).  Penton et al. (2004) have found broad Ly$\alpha$ lines as
well but in some cases have mandated that these be fit with multiple
components. Similarly, the automatic profile fitting algorithm used by
Dav\'{e} \& Tripp (2001) forced the broad features to be fitted with
multiple components. Some examples of broad Ly$\alpha$ lines are shown
in Figure~4a; see Richter et al. (2004) and Sembach et al. (2004b) for
additional plots of these broad \ion{H}{1} lines. Figure~4b shows the
$b-$values and \ion{H}{1} columns for all Ly$\alpha$ lines identified
by Richter et al. (2004) and Sembach et al. (2004b); this panel shows
a significant population of strikingly broad lines. If the broadening
of these lines is predominantly due to thermal motions, then the
implied temperature is in the WHIGM range, and their baryonic content
is substantial (similar to the \ion{O}{6} systems). It remains
possible that the breadth of these features is not due to the gas
temperature.  However, many of the broad Ly$\alpha$ lines are
well-detected (see, e.g., Figure 7 in Sembach et al. 2004b) and cannot
be instrument artifacts. While they could be continuum undulations
intrinsic to QSO spectrum, this appears unlikely. A more likely
explanation is that the broad Ly$\alpha$ lines are multiple, blended
components as assumed in some of the papers above.  Very high S/N
spectra would be valuable for testing this possibility: if the lines
are multiple blended features, departures from a simple Gaussian shape
could be evident in high S/N data. We note that some of the broad
Ly$\alpha$ lines presented in Sembach et al. (2004b) have good
signal-to-noise and appear to be well-described by a single, smooth
feature, i.e., there are no obvious indications of multiple components
at the current signal-to-noise level (see Figure 7 in Sembach et
al. 2004b). It would also be useful to compare the observed $b$
vs. N(\ion{H}{1}) (as in Figure~4b to predictions from hydrodynamical
simulations.

\begin{figure}
\plotone{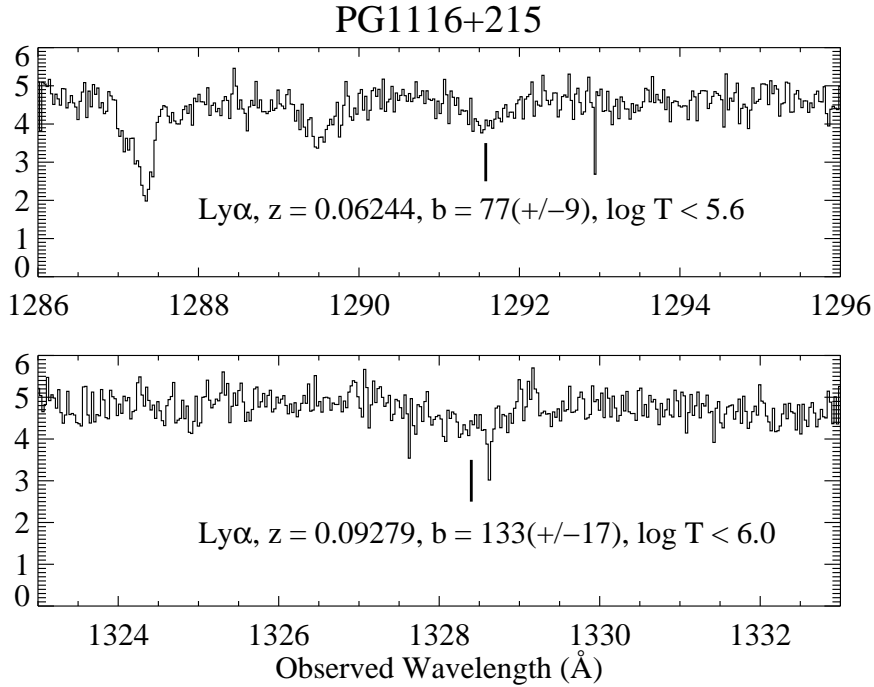}
\caption{(Figure 4a) Examples of two broad Ly$\alpha$ absorbers in the
spectrum of PG1116+215 (Sembach et al. 2004b). Note that the narrow
feature superimposed on the broad Ly$\alpha$ line in the lower panel
is a \ion{C}{1} line from the Milky Way ISM (many Galactic \ion{C}{1}
lines are detected in the PG1116+215 spectrum that corroborate this
identification; see Sembach et al. 2004b).\label{broadexamples}}
\end{figure}

\setcounter{figure}{3}
\begin{figure}
\plotone{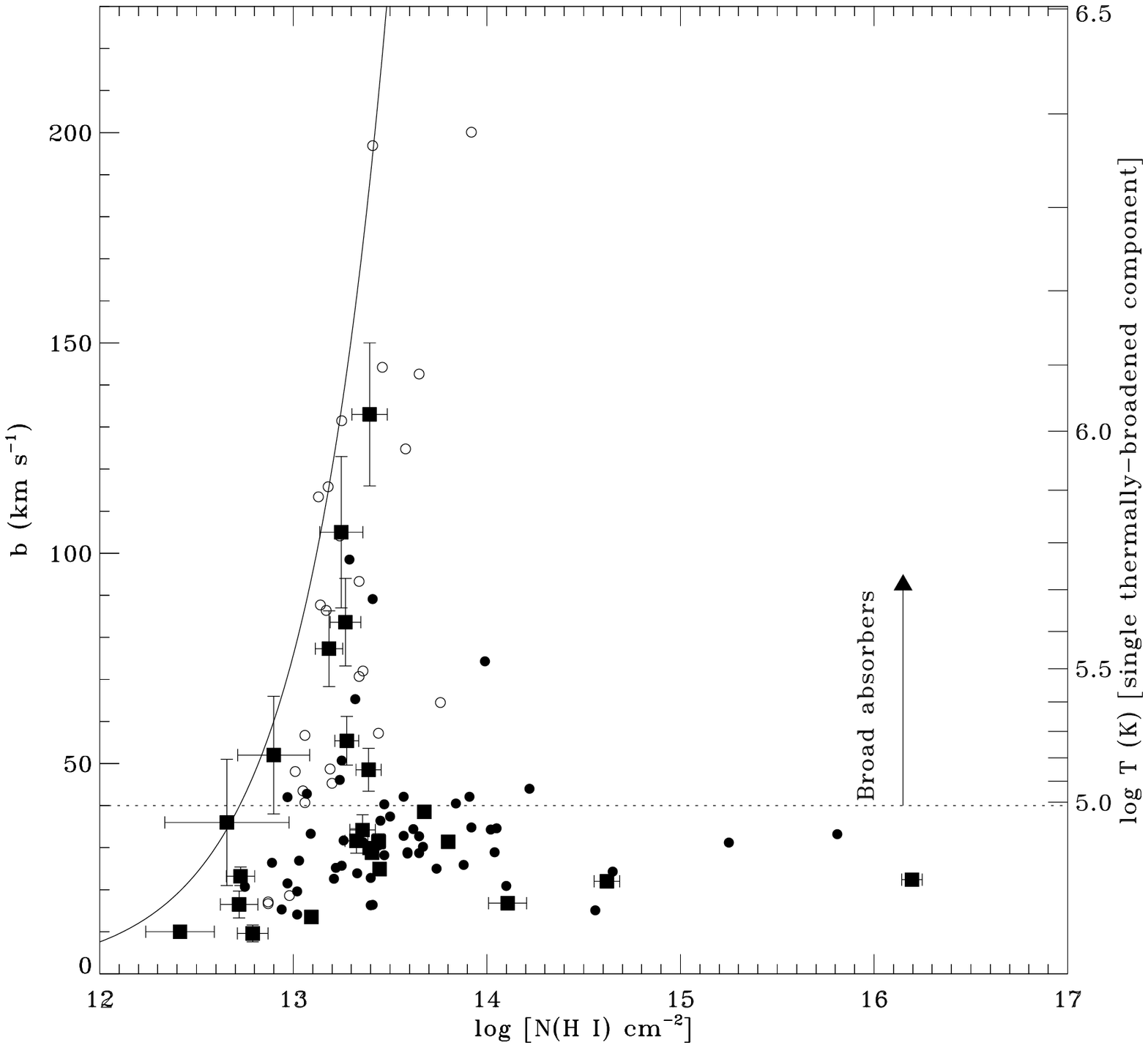}
\caption{(Figure 4b) Doppler parameters and \ion{H}{1} column
densities of all Ly$\alpha$ identified in the spectra of PG1259+593
and PG1116+215 by Richter et al. (2004) and Sembach et al. (2004b);
the solid line shows $b$ vs. $N$(\ion{H}{1}) for a Gaussian line with
10\% central optical depth. The right axis shows the temperature
implied by the line width if the lines are due to a single,
thermally-broadened component. In some cases, the smooth shape of the
Ly$\alpha$ profile supports this assumption (see Figure 7 in Sembach
et al. 2004b).\label{broadexamples2}}
\end{figure}

\section{Relationships with Galaxies}

\begin{figure}
\plotone{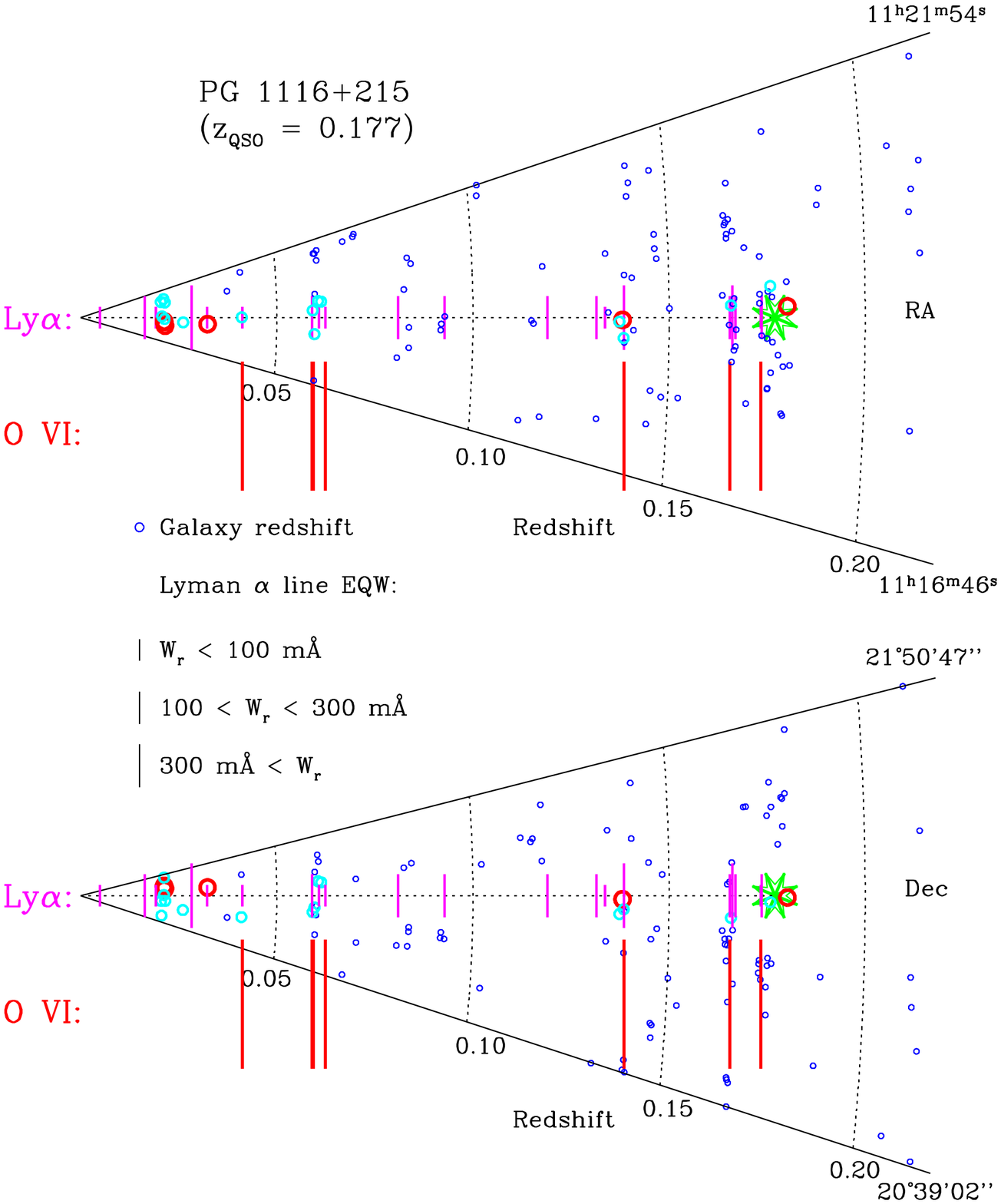}
\caption{Comparison of Ly$\alpha$ and \ion{O}{6} absorber redshifts
detected in the spectrum of PG1116+215 by Sembach et al. (2004b) to
galaxy redshift measurements from the WIYN survey of Tripp, Lu, \&
Savage (1998). The upper portion shows the RA wedge and the lower
shows Dec. In each wedge, Ly$\alpha$ line redshifts are indicated with
vertical lines overplotted on the sight line while \ion{O}{6} line
redshifts are shown with long lines offset from the sight
line. Galaxies are shown with circles, and the size of the circle
indicates proximity to the sight line: the largest circles mark
galaxies with projected distances ($\rho$) less than 500 kpc from the
line of sight; itermediate-size circles represent galaxies at $500 <
\rho < 1000$ kpc, and small circles show galaxies at $\rho >$ 1
Mpc.\label{galaxies}}
\end{figure}

Many papers have investigated the relationships between low$-z$
Ly$\alpha$ lines and nearby galaxies/structures. A review of this
literature is beyond the scope of this paper; a reasonable synopsis is
that the majority of the low$-z$ Ly$\alpha$ lines are strongly
correlated with galaxies but a few are found in voids (e.g., Penton et
al. 2000b; McLin et al. 2002), but despite the strong correlation, the
detailed nature of the Ly$\alpha$-galaxy relationship is not yet
well-understood. The Ly$\alpha$ lines almost certainly have a variety
of origins including (1) bound interstellar gas in the halo (or even
disk) of individual galaxies, (2) tidally stripped debris, (3)
galactic wind ejecta, and (4) truly intergalactic gas far from
galaxies.

Our understanding of the \ion{O}{6} absorber-galaxy connection is even
more immature, but already it is clear that the \ion{O}{6} systems are
strongly correlated with galaxies as well.  Several studies have
revealed galaxies within a few hundred kpc from \ion{O}{6} systems
(Savage, Tripp, \& Lu 1998; Tripp et al. 2000; Tripp \& Savage 2000;
Chen \& Prochaska 2000; Tripp et al. 2001; Savage et al. 2002; Shull
et al. 2003; Sembach et al. 2004b; Tumlinson et al. 2004). As an
example, Figure~\ref{galaxies} compares the locations of galaxies
found in the redshift survey of Tripp et al. (1998) to the redshifts
of \ion{O}{6} and Ly$\alpha$ lines detected by Sembach et
al. (2004b). Visual inspection of Figure~\ref{galaxies} suggests that
the \ion{O}{6} lines are correlated with the galaxies (see also Figure
21 in Sembach et al. 2004b for a striking visual comparison), and this
is corroborated by statistical tests: Sembach et al. (2004b) find that
the probability that the \ion{O}{6} absorbers are randomly distributed
with respect to the galaxies shown in Figure~\ref{galaxies} is $1.3
\times 10^{-4}$ to $2.0 \times 10^{-4}$ (depending on assumptions
made).  With the advent of powerful multiobject spectrographs on 8-10m
telescopes, future studies of the nature of \ion{O}{6} absorbers (as
well as other classes of absorption lines) and their connections with
galaxies hold great promise for understanding the role and
implications of these absorbers for galaxy evolution and cosmology.

\acknowledgements This research was supported in part by NASA LTSA
grant NNG04GG73G.  Many of the STIS observations were obtained under
the auspices of {\it HST} program 9184, with financial support through
NASA grant HST-GO-9184.08-A from the Space Telescope Science
Institute.





\end{document}